\documentclass[rnote]{aa}  
\usepackage{graphicx,natbib,ulem}
\usepackage{txfonts}
\usepackage{color}
\begin{document}
  \titlerunning{The wind of SS 433}
\authorrunning{ Bowler}
   \title{SS 433: The wiggle of the wind}

   \subtitle{}

   \author{M. G.\ Bowler \inst{}}

   \offprints{M. G. Bowler \\   \email{m.bowler1@physics.ox.ac.uk}}
   \institute{University of Oxford, Department of Physics, Keble Road,
              Oxford, OX1 3RH, UK}
   \date{Received 22 July 2011/ Accepted 12 September 2011}

 
  \abstract
   { The brilliant Balmer H$\alpha$ line in the stationary emission spectrum of the Galactic microquasar SS 433 has a broad component ($\sim$ 1000 km s$^{-1}$). This is formed in the wind blowing from the accretion disk of the compact object, which orbits the centre of mass of the binary at $\sim$ 175 km s$^{-1}$ with a 13-day period. The centroid of the H$\alpha$ emission line formed in the wind has an imperfect memory of the motion of its source.} 
  {The aim is to understand how this emission line is left with a blurred memory of the motion of the source of the wind.}
   {We analysed stationary H$\alpha$ spectra, taken almost
     nightly over two orbital periods of the binary system.}  
  { The loss of memory by emission lines from the wind is easily understood quantitatively, if any parcel of wind that has been newly conditioned to radiate in H$\alpha$ continues its emission for a period of days. If the decay is exponential, a mean lifetime of about two days is all that is required. This timescale is very close to what is required for the decay of H$\alpha$ from the putative circumbinary disk to be responsible for the narrow components of the H$\alpha$ emission line. It is also close to the timescales observed for the individual bolides that radiate H$\alpha$ in the relativistic jets.}
{ The structure of the stationary H$\alpha$ emission line is now well understood. The broad component is undoubtedly formed in the wind from the accretion disk and the two narrow components carry no such signature. Their properties are entirely consistent with an origin in a circumbinary disk. The characteristic lifetime of H$\alpha$ emitting material is a few days, as observed for the relativistic bolides in the jets.}


   \maketitle
%

\section{Introduction}

We consider a very luminous star blowing a dense wind and a particular shell within that wind. It is accelerated rapidly to the terminal velocity, perhaps several thousand km s$^{-1}$. When close to the terminal velocity that expanding shell is effectively detached from the star and thereafter continues to expand with its centre following a straight-line trajectory, the velocity vector of which is equal to the velocity vector of the source at the time of detachment. If the very luminous source is a member of a binary system, the trajectory of the centre of a shell of wind does not deviate as the star continues in its closed orbit. Thus if emission lines are formed in the shell after the time of detachment, the mean Doppler shift of the shell spectrum will reflect the velocity vector of the source at the time of detachment rather than at the time of formation of the emission lines. If any given shell glows for a time lasting a significant fraction of the binary orbit, at any one moment the centre of any emission line will reflect the velocity of the source, but delayed and averaged over part of the orbit. The centre of an emission line formed in the wind will (for an approximately circular orbit seen approximately edge on) execute a sinusoidal oscillation with a velocity amplitude smaller than that of a line formed in the atmosphere co-moving with the star. This scenario is speculative, but might be relevant to binary systems containing Wolf-Rayet stars or objects with at least some of their observational characteristics. One such object is the Galactic microquasar SS 433, and in this note I suggest that the emission spectrum associated with the wind from the accretion disk exhibits just this behaviour.

SS 433 is very luminous and famous for its continual ejection
of plasma in two opposite jets at approximately one quarter the speed
of light. The system is a 13-day binary ( probably powered by supercritcal accretion by the compact member from its companion), and the orbital speed of the compact object is fairly well established as $\sim$ 175 km s$^{-1}$. Despite the fame of the emission lines from the relativistic jets, the so-called stationary emission lines are much more intense, in particular the brilliant H$\alpha$ line. Most aspects of the system are reviewed in Fabrika (2004). The stationary lines in fact are not quite stationary, rather  they exhibit sinusoidal Doppler shifts with a 13-day period; the oscillation of H$\beta$ and He I stationary emission lines first demonstrated that SS 433 is a binary (Crampton, Cowley \& Hutchings 1980). The velocity amplitudes of the Balmer series and He I lines, when treated as though they come from a single source, are very roughly 70 km s$^{-1}$ and do not provide a direct measure of the orbital speeds of either the compact object or its companion, because they are redshifted most when the compact object is very close to eclipse by the companion (orbital phase 0), rather than a quarter of a period earlier when the compact object is receding fastest from us (orbital phase 0.75); see for example Crampton \& Hutchings (1981) and Gies et al (2002). This phasing of the emission lines relative to the photometric ephemeris naturally suggests that the Balmer and He I emission lines are formed in an accretion stream passing from the donor to the companion. However, Gies et al (2002), evidently influenced by the extreme brilliance of the stationary H$\alpha$, proposes that these emission lines are formed in the wind from the accretion disk, but at distances comparable to or greater than the dimensions of the binary system. In that paper it is suggested that interference with the wind from the disk by the companion, or indeed by a wind from the companion, creates a void in the disk wind. This void would enfeeble the blue part of the spectrum when the companion is between us and the compact object (and conversely enfeeble the red half a period later). The origins of the stationary emission lines were later clarified by analysis of the stationary H$\alpha$ line, in terms of a superposition of several Gaussian profiles (Blundell, Bowler \& Schmidtobreick 2008), exploiting a remarkable sequence of spectra taken nightly over several orbits (see Schmidtobreick \& Blundell 2006).

 \begin{figure}[htbp]
\begin{center}
   \includegraphics[width=15cm]{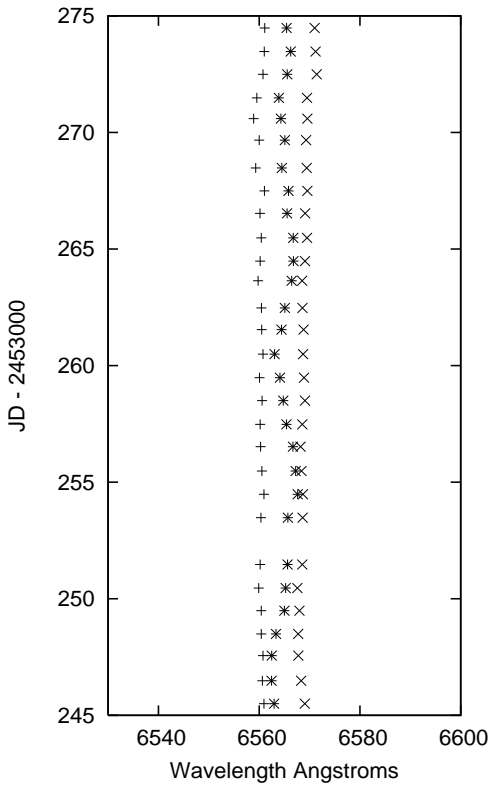} 
\caption{ Data from Fig.1 of Blundell, Bowler \& Schmidtobreick (2008) for the wavelengths of the centroids of three Gaussians fitted to the H$\alpha$ spectra. The blueshifted narrow Gaussian is denoted by +, the redshifted by x, and the broad Gaussian (the H$\alpha$ profile of the wind) by $*$. The centre of the wind profile wanders back and forth between the railroad tracks, which have been attributed to a circumbinary disk. Orbital phase 0 (primary eclipse) occurs at JD +255.
 }
\label{fig:railroad}
\end{center}
\end{figure}

\section{H$\alpha$, He I, and three-component analysis}

The spectra used in Blundell, Bowler \& Schmidtobreick (2008) are displayed as a montage in Fig.2 of Schmidtobreick \& Blundell (2006). That data set extends from JD 2453245 until JD 2453321, and in Blundell, Bowler \& Schmidtobreick (2008) the spectra were analysed up to JD +274, after which there was something of a hiatus, followed by an optical outburst and a radio flare (Blundell, Schmidtobreick \& Trushkin 2011). Up to JD +274, most H$\alpha$ spectra were adequately fitted with three Gaussians, each characterised by centroid and a width parameter taken as the standard deviation of the Gaussian, $\sigma$. Two were narrow, having $\sigma$ of only a few $\AA$. The third component was much broader, with $\sigma$ up to 20 $\AA$. This third component was identified with an origin in the wind above the accretion disk, and it is the subject of the present paper. Fig.1 of Blundell, Bowler \& Schmidtobreick (2008) displays the centroids of these fitted Gaussians as a function of time, in the same sequence as the montage in Fig.2 of Schmidtobreick \& Blundell (2006). Because of their importance, I have reproduced these data in Fig.1. This figure shows that the two narrow Gaussian components of H$\alpha$ have centroids that scarcely move over more than two orbits, and the broad component oscillates in Doppler shift with a period of 13 days. It is most redshifted close to orbital phase 0, at primary eclipse. This might suggest that this component is formed in an accretion stream rather than in the wind. The two narrow components were identified with  the inner rim of a circumbinary disk in Blundell, Bowler \& Schmidtobreick (2008).

  A very similar plot has been obtained from a set of H$\alpha$ spectra taken between 2004 and 2008 at the Purple Mountain Observatory. The original data are to be found in Fig.2 of Li \& Yan (2010) and the motions of fitted centroids as a function of orbital phase are displayed in Fig.2 of Bowler (2011a). The same features are apparent.
  
  In Fig.2 I show the variation of Doppler speed for the centre of the H$\alpha$ broad component (Blundell, Bowler \& Schmidtobreick 2008) and above it the same for the broader component of the He I 6678 \AA\ line, obtained from my fits reported in Bowler (2011b). These broad components have centroids that oscillate with a 13-day period and with velocity amplitude $\sim$110 km s$^{-1}$. They are most redshifted a little before orbital phase 0 and most blueshifted a little before 0.5. Lines produced in an atmosphere co-moving with the compact object would be most redshifted at orbital phase 0.75 and most blueshifted at 0.25 - the broad lines associated with the wind lag the motion of the compact object by $\sim$ 0.2 of a period and have a reduced orbital Doppler amplitude.
  
  The identification of the broad component of the H$\alpha$ emission line with an origin in the wind from the disk is discussed in Blundell, Bowler \& Schmidtobreick (2008). In brief, the width parameter $\sigma$ drops smoothly from approximately 20 \AA\ near JD +245 (precession phase approximately 0; accretion disk wide open) to 10 \AA\ near JD +274 as the accretion disk comes closer to edge on. In addition, this measure of the line of sight wind speed follows the nodding motion of the accretion disk, as inferred from the Doppler shifts of the relativistic jets (Blundell, Bowler \& Schmidtobreick 2007). The synthesis of absorption line studies of the wind (Fabrika 1997, 2004) shows that the line of sight expansion speed varies as approximately the square of the cosine of the angle between the jet axis and the line of sight, reaching 1600 km s$^{-1}$ for 60$^{\circ}$. Thus these observations have established that the source of this wind line is rooted in the accretion disk. In Blundell, Bowler and Schmidtobreick (2008) the authors cautiously observed that despite this, the motion of the centroid of the wind should not be taken as a measure of the orbital velocity of the compact object. If a given parcel of wind continues emitting H$\alpha$ over a period of several days, the data displayed in Fig.2 are reconciled with an origin in the tilting and nodding accretion disk, according to the scenario sketched in the introduction.

\begin{figure}[htbp]
\begin{center}
  \includegraphics[width=9cm,trim=0 0 0 80]{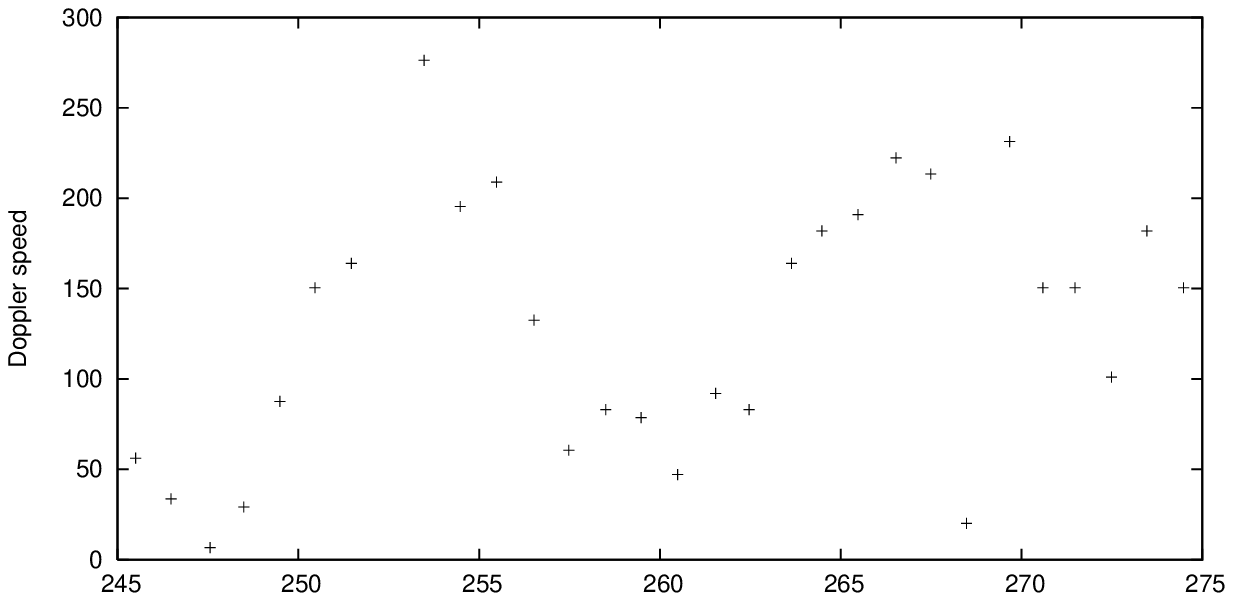}
   \includegraphics[width=9cm,trim=0 0 0 50]{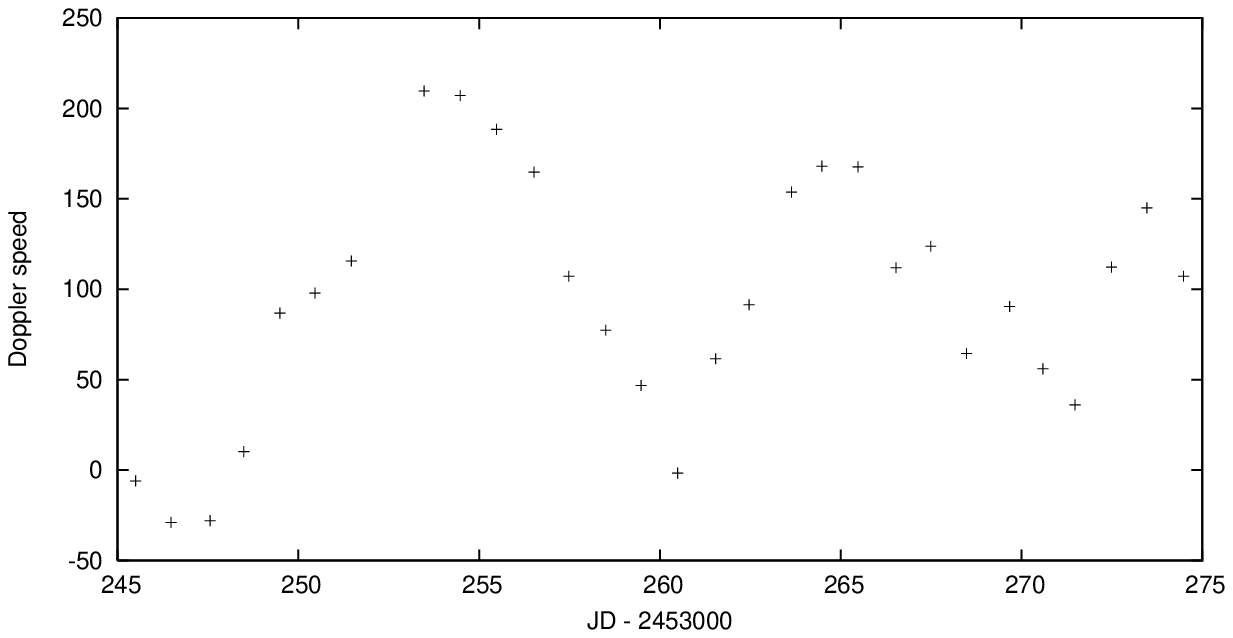} 
   \caption{ The line of sight recession speed of the centroid of the broadest Gaussian component (the wind) as a function of date of observation. The upper panel is for He I 6678 \AA\ and the lower panel for H$\alpha$. Orbital phase 0 (primary eclipse) occurs at JD +255 and JD +268.}
\label{fig:timesequence}
\end{center}
\end{figure}

 \section{Quantitative considerations}
 
 Absorption line studies have shown that the wind from SS 433 is slow in the plane of the accretion disk, but that the speed increases rapidly as the angle to the jet axis decreases. The results are summarised in Fabrika (1997, 2004) in the form
 
 \begin{equation}
 V_w = 8000 \cos^2\alpha + 150
 \end{equation}
where $V_w$ is the velocity (in km s$^{-1}$) of the gas flowing out from the disk as a function of the polar angle $\alpha$. The width $\sigma$ of the broad H$\alpha$ component in the data of Blundell, Bowler \& Schmidtobreick (2008) has a value of $\sim$20 \AA\ early on, when the angle $\chi$ between the jet axis and the line of sight has a value of $\sim$ 60$^{\circ}$, declining to 10 \AA\ by JD +274, when  $\chi$ has a value of $\sim$ 70$^{\circ}$. These widths correspond to speeds of 900 and 450 km s$^{-1}$, respectively, and the data are represented fairly well by the relation

\begin{equation}
s_w = 3000 \cos^2\chi
\end{equation}
where $s_w$ is the width $\sigma$ translated into speed in km s$^{-1}$. The speeds in the wings of the wind H$\alpha$ line are even greater. Thus typical wind speeds are very roughly an order of magnitude greater than the orbital speed of the compact object, and a parcel of wind covers a distance of roughly 1.5 $10^{8}$ km in a single day. This distance is several times larger than the semi-major axis of the binary system and because the fast wind is predominantly perpendicular to the accretion disk (see also Perez \& Blundell 2010) it tends to be out of the plane of the orbit.

It remains to make a quantitative estimate of the effect of any given region of the wind emitting over a period of perhaps several days. I approximate the line of sight to the orbital plane (it actually makes an angle of $\sim$12$^{\circ}$) and suppose a circular orbit, so that the recessional velocity of a source co-moving with the compact object would be given by

\begin{equation}
v_r = -v_x \sin \omega t
\end{equation}
where $\omega$ is given by $2\pi /P$. Thus when $t$ is one quarter of the period $P$ after primary eclipse, the compact object is approaching with speed $v_x$ and is receding with speed $v_x$ a quarter of a period before primary eclipse.

I now assume that a shell of wind lights up suddenly (say in H$\alpha$) at time $t$ but became detached a time $s$ earlier, the centroid moving with speed $v_x$ tangential to the orbit. For a delay $s$ of $0.25P$, the most redshifted centroid is observed when $t=P$ rather than $0.75P$. The phasing of the wind centroid relative to the photometric ephemeris only requires a delay of $\sim$2 days. The final step is to suppose that the H$\alpha$ emission dies away over a timescale of several days. This will affect the phase and also makes an average of the line-of-sight velocity of the H$\alpha$ centroid, thereby reducing the amplitude of $v_r$ below $v_x$. This may be calculated by specifying some emission function of $s$ with a delay parameter $\tau$ and a duration parameter $T$; $f(s; \tau ,T)$. At time $t$, the centroid of the shell detached a time $s$ earlier has recessional velocity given by

\begin{equation}
v_r(t,s) = -v_x\sin\omega (t-s).
\end{equation}

The value perceived at time $t$ is obtained by averaging over all $s$, using $f(s;\tau ,T)$ as the weight. The important point is that for durations of a few days the average over $s$, $<v_r(t)>$, represents very well the data in both amplitude and phase. For the purpose of illustration I have used two different functions for the emission factor $f(s)$. In the first case I supposed a rectangular profile as a function of $s$, with a duration of $T$. This switches on at $s= \tau - T/2$ and switches off at $s= \tau + T/2$, it being supposed that $\tau$ exceeds $T/2$. The weighted average is then

\begin{equation}
<v_r(t)> = -v_x\sin\omega (t-\tau)\sin (\omega T/2)/(\omega T/2).
\end{equation}

This is not a realistic form, but it makes the point and the structure is easy to visualise. For the particular case of the parameter $\tau = T/2$ (ignition immediately on launch) and duration time $T=P/2$, the centroid of the wind is most redshifted at orbital phase 0, one quarter of a period late, and the amplitude is $v_x$sinc$(\pi /2)$, which for $v_x$ 175 km s$^{-1}$ is 111 km s$^{-1}$. These results are very close to the behaviour of the data. The parameters $T$ and $\tau$ might exhibit some fluctuation with time - winds can be gusty.

A probably more realistic form is to suppose exponential decay of the emission factor after the initial light up. In this case the duration parameter $T$ is the decay time of the exponential, and there is a delay $\tau$ between launch and ignition. The weighted average recession is now

\begin{equation}
<v_r(t)> = \frac{v_x}{1 + (\omega T)^2} [-\sin\omega (t-\tau) + \omega T\cos\omega (t-\tau)] .
\end{equation}

For the simple case of $\omega T=1$, the decay time $T$ is about two days, the amplitude of the oscillation is $v_x/\sqrt 2$ (124 km s$^{-1}$) and the recession velocity is greatest at orbital phase 0.875 for the case of instant ignition, $\tau =0$. Thus these simple models have demonstrated that both the amplitude and the phase of the centroid of the broad component of H$\alpha$, relative to the photometric ephemeris, are easily understood in terms of a wind that becomes detached from the orbiting source and decays away in H$\alpha$ over a few days. The amplitude and phase of the wind centroid is reconciled with the way in which the line-of-sight wind speed varies with the nodding of the disk (Blundell, Bowler \& Schmidtobreick 2008).

\section{Two observations from more violent times}

The period from JD +245 to + 274 was a period of calm before the storm. The persistent pattern in H$\alpha$ of two narrow components with a broad wind swinging between them was overwhelmed by an optical outburst commencing around JD +290, followed by a radio flare. Between JD +291 and +295 the speed of the wind, as measured by the width $\sigma$ of the broad component, rose from $\sim$600 km s$^{-1}$ to $\sim$1200 km s$^{-1}$ and then remained at the higher figure. The history is presented in some detail in Blundell, Schmidtobreick \& Trushkin (2011) - see Fig.8 therein. In the lower panel of Fig.3, I show the motion of the centroid of the wind in H$\alpha$ over the period JD +294 to +310, as displayed in Fig.4 (c) of Blundell, Schmidtobreick \& Trushkin (2011). The amplitude is certainly consistent with the orbital speed of the compact object, and despite the paucity of data, the first minimum (blueshift) occurs at approximately JD +298, an orbital phase only marginally above 0.25. Similarly, the following maximum redshift occurs at $\sim$ JD +304, marginally above orbital phase 0.75. These data seem to have a better memory of the motion of the source of the wind than the slower winds in quieter times; in terms of the exponential decay model the data suggest a decay time of perhaps one day or less. 

Over this same period many of the He I spectra are rendered unusable by the proximity of the moving H$\alpha$ lines of the jets, but after JD +294 the O I triplet commencing at 7772 \AA\  exhibits a pronounced P Cygni absorption trough. These O I spectra may also be found in Fig.2 of Schmidtobreick \& Blundell (2006), and from that figure I have extracted the motion of the deepest point of the trough. This is shown in the upper panel of Fig.3. The pattern is erratic but does show some memory of the orbital motion of the compact object, with an amplitude of $\sim$ 100 km s$^{-1}$, with minimum at orbital phase $\sim$ 0.5 and maximum at JD +306, orbital phase close to 1. These data suggest that the O I 7772 \AA\ line is formed in the slow equatorial wind from the disk, with a speed $\sim$150 km s$^{-1}$ and that (absorption) decay times $\sim$2 days are appropriate. [ The O I 7772 \AA\ line is not accessible during the period JD +245 to +274 because of the redshifted moving H$\alpha$ line.]

The notion of a detached wind affecting the magnitude and phase of Doppler shifts may also be relevant to certain absorption spectra in the blue. These have features similar to mid-A supergiants and are probably formed in the wind (Barnes et al 2006).

\begin{figure}[htbp]
\begin{center}
  \includegraphics[width=9cm,trim=0 0 0 80]{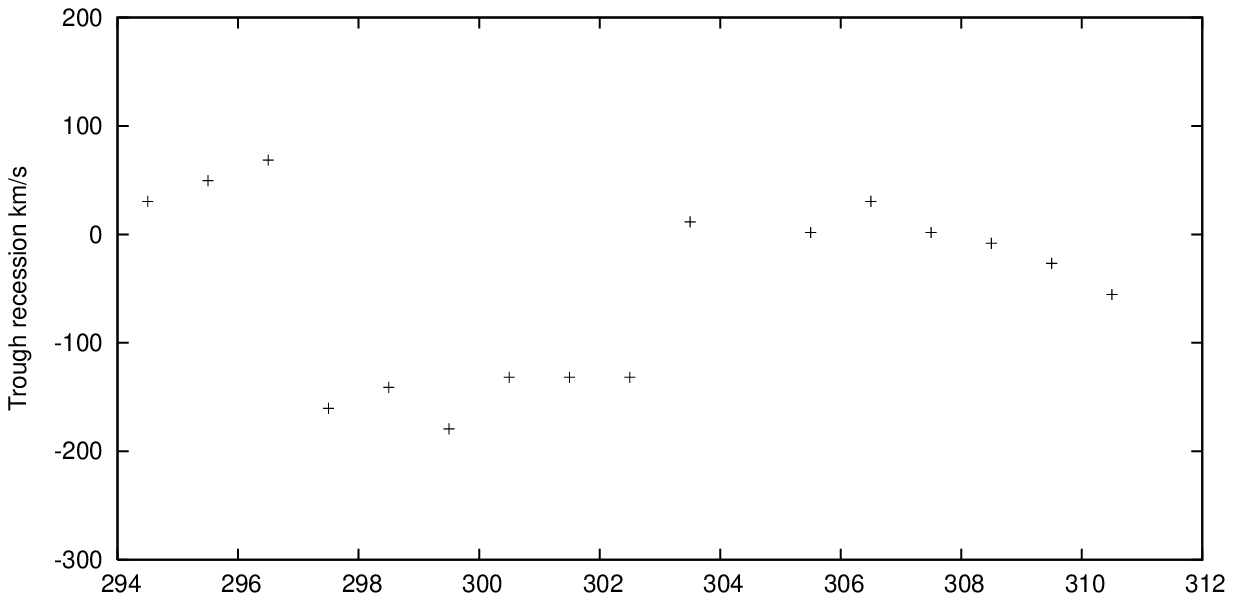}
   \includegraphics[width=9cm,trim=0 0 0 50]{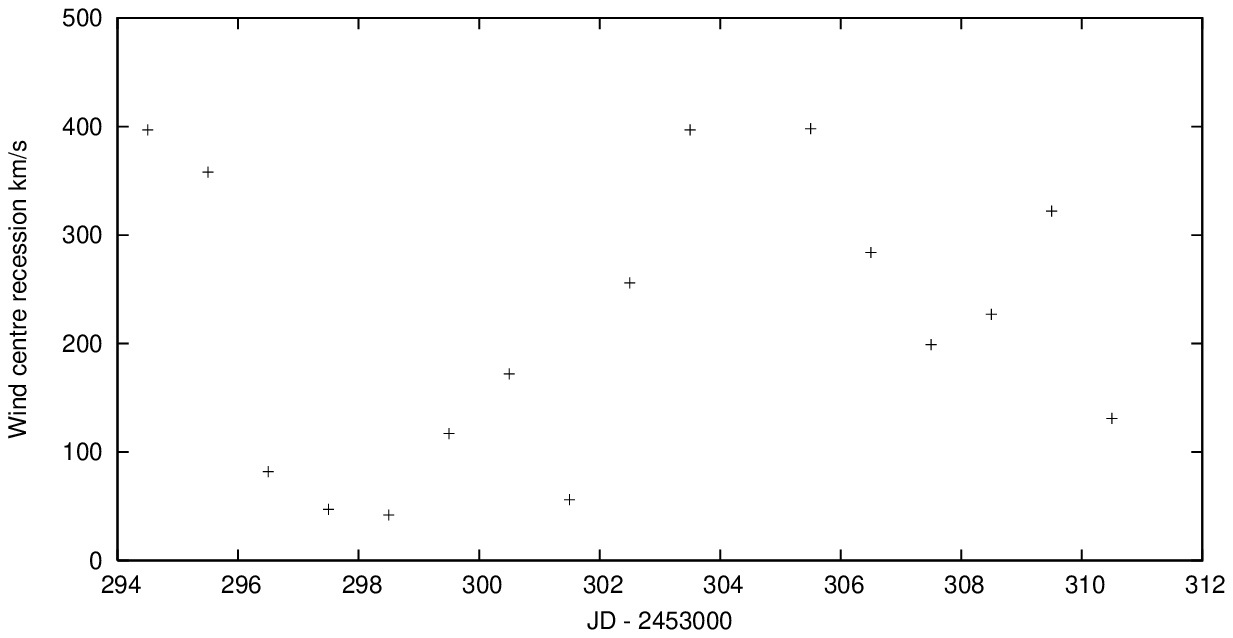}
   \caption{ Lower panel:  the motion of the centroid of the fast H$\alpha$ wind during the outburst episode. Upper panel: the motion of the blueshifted absorption trough in O I 7772 \AA\ .}
\label{fig:timesequence}
\end{center}
\end{figure}

  \section{Discussion}
  
  The broad component of the stationary H$\alpha$ line was identified as formed in a wind from the accretion disk of SS 433 on the basis of the line of sight speed varying with the nodding of the disk (Blundell, Bowler \& Schmidtobreick 2008). The amplitude and phase of the motion of the centroid of this component have now been shown to be entirely consistent with this origin, provided the emission decay time within the wind is a few days. The broad component of the H$\alpha$ is indeed formed in the wind on a scale comparable to or greater than that of the binary system (Gies et al 2002). In contrast, the two narrow components show no sign of an origin within the wind; indeed, the narrow components of both H$\alpha$ and He I are not consistent with such an origin but are well described in terms of emission from a fairly close circumbinary disk, stimulated by the intense radiation from the vicinity of the compact object (Bowler 2010, 2011b). It is interesting that in that circumbinary disk model the decay time for H$\alpha$ emission  must again be of the order of a few days. Thus there are two phenomena, not directly related, that are described very well by the phenomenological assumption that H$\alpha$ emission from dense plasma has, following ignition, a decay time of the order of days. There remains the question of by what mechanisms H$\alpha$ radiation is sustained for several days after ignition. That at least one mechanism exists enabling clumps of plasma to exhibit H$\alpha$ decay times of a few days is certain: the individual bolides ejected at about one quarter the speed of light in the jets for which SS 433 is famous are visible in H$\alpha$ for several days ( see for example Vermeulen et al 1993, Gies et al 2002, Blundell, Bowler \& Schmidtobreick 2007 ).

\end{document}